\documentclass[conference]{IEEEtran}
\IEEEoverridecommandlockouts
\usepackage{cite}
\usepackage{amsmath,amssymb,amsfonts}
\usepackage{algorithmic}
\usepackage{graphicx}
\usepackage{subcaption}
\usepackage{textcomp}
\usepackage{fancyhdr}
\usepackage{lipsum}
\usepackage{xcolor}
\usepackage{array}
\def\BibTeX{{\rm B\kern-.05em{\sc i\kern-.025em b}\kern-.08em
    T\kern-.1667em\lower.7ex\hbox{E}\kern-.125emX}}
    
\usepackage{listings}
\usepackage{color}
\definecolor{dkgreen}{rgb}{0,0.6,0}
\definecolor{gray}{rgb}{0.5,0.5,0.5}
\definecolor{mauve}{rgb}{0.58,0,0.82}

\fancypagestyle{plain}{
  \fancyhf{}
  \fancyfoot[L]{This is a notice}

}
\usepackage{tikz}
\usetikzlibrary{calc}

\begin{document}

\title{
The Impact of Auto-Refactoring Code Smells on the Resource Utilization of Cloud Software
\\
}

\author{\IEEEauthorblockN{Asif Imran\IEEEauthorrefmark{1} and
Tevfik Kosar\IEEEauthorrefmark{2}}
\IEEEauthorblockA{Department of Computer Science and Engineering,
University at Buffalo\\
Amherst, New York 14260-1660, USA\\
Email: \IEEEauthorrefmark{1}asifimra@buffalo.edu,
\IEEEauthorrefmark{2}tkosar@buffalo.edu}}
\maketitle

\maketitle

\begin{tikzpicture}[remember picture, overlay]
\node at ($(current page.south) + (-5.85,0.65in)$) {\small DOI reference number: 10.18293/SEKE2020-138};
\end{tikzpicture}

\begin{abstract}
Cloud-based software-as-a-service (SaaS) have gained popularity due to their low cost and elasticity. However, like other software, SaaS applications suffer from code smells, which can drastically affect functionality and resource usage. Code smell is any design in the source code that indicates a deeper problem. The software community deploys automated refactoring to eliminate smells which can improve performance and also decrease the usage of critical resources. However, studies that analyze the impact of automatic refactoring smells in SaaS on resources such as CPU and memory have been conducted to a limited extent. Here, we aim to fill that gap and study the impact on resource usage of SaaS applications due to automatic refactoring of seven classic code smells: god class, feature envy, type checking, cyclic dependency, shotgun surgery, god method, and spaghetti code. We specified six real-life SaaS applications from Github called Zimbra, OneDataShare, GraphHopper, Hadoop, JENA, and JAMES which ran on Openstack cloud. Results show that refactoring smells by tools like JDeodrant and JSparrow have widely varying impacts on the CPU and memory consumption of the tested applications based on the type of smell refactored. We present the resource utilization impact of each smell and also discuss the potential reasons leading to that effect.

\end{abstract}

\begin{IEEEkeywords}
Code smells, automated refactoring, cloud resource utilization
\end{IEEEkeywords}

\section{Introduction}
Software as a Service (SaaS) running in cloud platforms has become increasingly popular, mainly due to their lower cost, high availability, and quality of service for the users. SaaS applications are becoming mainstream in the everyday lives of users who use them to conduct day-to-day business and personal software needs \cite{pham2017paas}. SaaS applications are rapidly capturing the market, and more service providers are migrating their software to cloud everyday \cite{armbrust2010view}. The critical advantage of SaaS is its capability to serve millions of users all around the world, harnessing the elasticity, reliability, and scalability of the underlying cloud platform.

Developing SaaS applications differ from desktop applications since those are required to perform on the cloud backbone where multiple software are competing for the compute cloud resources, mainly CPU and memory \cite{imran2014cloud}. Furthermore, the shorter time-to-market affects the development of those open-source software, which are designed to serve multiple users. Those applications are designed by many contributors who work together to solve common issues and regularly add new features. When open source SaaS is incorrectly coded, they can drain cloud resources like CPU and memory, thereby resulting in wastage of critical resources that are billed by cloud service providers \cite{imran2014cloud}. This results in technical debt and increases the cost of hosting the software in the cloud. We call the erroneous designs as code smells of SaaS.

Resource consuming code smells can degrade the performance of the SaaS application and increase the cost. Users will be demotivated to pay the increased cost and can move to less expensive services provided by the competitors. As a result, fixing those smells can improve the cost-efficiency of critical computing resources without sacrificing the quality of service. Hence, selectively refactoring these code smells would benefit the SaaS service providers as well as the customers of those services.
Previous research has focused on exploring the impact of code smells on energy consumption, such as battery usage in smartphones \cite{verdecchia2018empirical}. Also, the importance of detecting and eliminating smells that affect speedup during migration has been addressed \cite{wang2014platform}. However, there is a lack of empirical study which focuses on analyzing the impact of automatically refactoring smells in SaaS on CPU and memory consumption.   

This paper aims to fill the void by analyzing the effect of automatic refactoring of smells on resource utilization of SaaS applications running in the cloud. We conducted an initial search that included 84 repositories and filtered those to match the required pre-specific criteria of this research. We selected seven code smells, which we detected in the selected software using two popular tools,  JDeodrant~\cite{fokaefs2011jdeodorant} and JSparrow~\cite{jsparrow}, which have capabilities of detecting classic code smells and applying automatic refactoring on those. Out of the seven smells, JDeodrant detected and refactored {\em god class, feature envy,} and {\em type checking} whereas JSparrow could detect and refactor four smells namely {\em cyclic dependency, shotgun surgery, god method,} and {\em spaghetti code}. We strongly believe our research efforts will help to identify the critical importance of refactoring specific code smells in cloud-based software and their impact on the utilization of precious cloud resources.

The rest of the paper is organized as follows: Section II describes the methodology of our study, Section III presents the results of the experiments and summary of our findings, Section IV discusses the related work in this area, and Section V concludes the paper.


\begin{table*}[]
{\renewcommand{\arraystretch}{3}
\small
    \centering
    
    \begin{tabular}{|p{1.3cm}|p{6cm}|p{6cm}|p{3cm}|} 
    \hline
    Smell   & \parbox{6cm}{Primary Causes} & Impact on Software & Refactoring technique \\
    \hline
    \parbox{1cm}{cyclic dependency}&\parbox{6cm}{-Violation of acyclic modularization \cite{sarkar2012measuring}\\-Misplaced elements\\-Two packages are dependent on each other\\-Lack of encapsulation}&\parbox{6cm}{-Difficult to maintain\\-Single method called for multiple tasks\\-Has ripple effect on other abstractions \cite{sarkar2012measuring}}&\parbox{3cm}{-Encapsulate all packages in a cycle and assign to single team}
    \\
    \hline
    god method&\parbox{6cm}{-Many activities in a single method \cite{Fowler}\\-Tangled code\\-Long methods with multiple responsibilities}&\parbox{6cm}{-Memory leak\\-Multiple calls to same function \cite{lanza2007object}\\-Difficult co-ordination of packages}&\parbox{3cm}{-Extract method\\-Replace method with method object}\\
    \hline
    type checking&\parbox{6cm}{-Using complex variation of an algorithm requiring execution based on the value of an attribute \cite{tsantalis2008jdeodorant}\\-Objects in class come from different workers}&\parbox{6cm}{-Abuse of type casting\\-Redundant code in a method or class\\-Less flexible code}&\parbox{3cm}{-Replace "instanceof" from code\\-Strategy pattern}\\
    \hline
    spaghetti code&\parbox{6cm}{-Convoluted code\\-Continuous addition of new code and no removal of obsolete ones  \cite{abbes2011empirical}\\-Procedural code design}&\parbox{6cm}{-Difficult to understand code \cite{abbes2011empirical}\\-Lack of well-articulated code}&\parbox{3cm}{-Replace procedural code segments with object oriented design}\\
    \hline
    feature envy &\parbox{6cm}{-Accessing data of another object often \cite{fokaefs2011jdeodorant}\\-Occurs when fields are moved to data class\\-Data defined in class A, however operations defined in class B}&\parbox{6cm}{-High volume of requests to access a class and its objects\\-Numerous read and write to remote object}&\parbox{3cm}{-Move method\\-Extract method}\\
    \hline
    shotgun surgery &\parbox{6cm}{-Single behavior defined across multiple classes\cite{Fowler}}&\parbox{6cm}{-Requires multiple changes in different locations (e.g., multiple files) of the code in order to make a single modification \cite{Fowler}\\-Existing behavior in multiple classes}&\parbox{3cm}{Inline class}\\
    \hline
    god class &\parbox{6cm}{-Class aims to do many activities \cite{Fowler}\\-Large number of instance variables declared in one class}&\parbox{6cm}{-Difficult to manage multiple functionalities\\-Difficult to understand code}&\parbox{3cm}{-Extract class\\-Extract Interface}\\
    \hline

 \end{tabular}
    \caption{Characteristics and applied refactoring properties of software smells}
    \label{tab00}}
\end{table*}

\section{Experimental Process}
This section describes the goals of this study and the research questions answered, the selection of code smells, and the methodology of experimental and evaluation processes.
\subsection{Identified Research Questions}
The main goal is to determine whether automatic refactoring of smells in cloud-based software impact the resource consumption in terms of CPU and memory. The motivation is derived from the fact that cloud resources are in high demand, and any unnecessary resource usage will incur unnecessary costs. More specifically, we want to determine the refactoring of which smells will result in an increase or reduction of CPU and Memory consumption in the cloud. Based on this goal, we determine the following research questions:
\begin{itemize}
\item \textbf{RQ1.} How does auto-refactoring of code smells in cloud-based software affect CPU utilization?
\item \textbf{RQ2.} How does auto-refactoring of code smells in cloud-based software affect memory utilization?
\end{itemize}

\subsection{Selected Code Smells}
A code smell is a software behavior that is indicative of more profound quality issues~\cite{Fowler}. We selected the aforementioned seven smells mainly because of the following reasons. Primarily, those smells are studied popularly by software community, and they are considered as classic smells \cite{Fowler}. Secondly, several tools can detect the code smells; however, few tools are available, which can automatically refactor those. Our selected smells could be automatically detected by the refactoring tools we used in this study. Table \ref{tab00} provides a summary of the smells which includes the causes, impact on software, and the refactoring policy applied by the tools. 
\subsection{Study Methodology}
To eliminate biases in our study, we executed each of the apps 24 times and took the mean of the data. To analyze the effect of a given smell, first, we ran the smelly code 3 times. Then we refactored a smell and ran the software 3 times more, hence for one software, we had (7*3=21)+3=24 runs. Running each software 24 times resulted in a total of 144 runs. This required a significant time frame.
\begin{figure*}
    \centering
    \includegraphics[width=18cm, height=3.7cm]{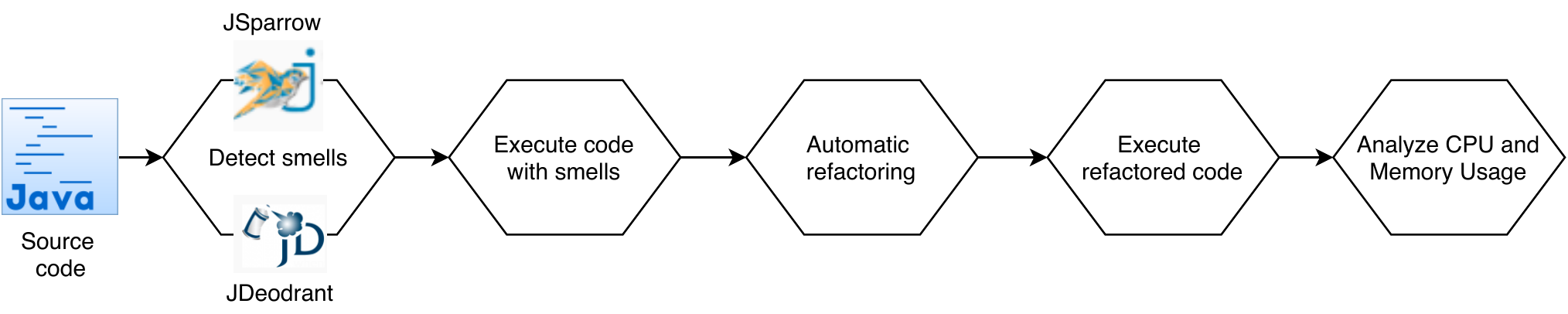}
    \caption{Automatic detection and refactoring of code smells.}
    \label{fig1}
\end{figure*}

We conducted the experiments in $OpenStack$ cloud servers, which had 32 GB RAM, 8 core processors, and 2 TB persistent storage \cite{imran2016web}. We set up the cloud and ran the six software in VM instances, which were created using \textit{kernel virtual machine (kvm)}. Every time we refactored, we executed the software in a new instance and deleted all existing data to minimize the impact of caching. We did not run any other cloud VM instance with other services to minimize the effect of external entities. Following this mechanism allowed us to erase all existing data before the software was run. As a result, it helped us achieve the same initial condition for each experimental run. Next, we followed a methodology for refactoring the selected software, which is shown in Figure \ref{fig1} and discussed here. 
\begin{itemize}
    \item \textbf{Smell detection and provenance:} As a first step, we applied \textit{JDeodrant} to detect the three code smells ({\em god class, feature envy,} and {\em type checking}), followed by \textit{JSparrow} to detect four remaining smells. Since both \textit{JDeodrant} and \textit{JSparrow} have plugins for \textit{Eclipse IDE}, we imported the source codes in to \textit{Eclipse} and built those. After detection of the smells, we preserved provenance of the packages, classes, and methods, which were affected by the smells. Overall, \textit{JDeodrant} and \textit{JSparrow} detected 744 instances of the seven analyzed smells in the six software. The reason for using two tools is to include significant number of smells in the study. None of the tools could single-handedly detect and refactor all the smells considered here. 
    \item\textbf{Executing smelly code:} Next, we executed the codes in the cloud. We ran the code three times, each time with the same workload, and collected the CPU and memory utilization during each clock cycle using scripts in $OpenStack$ \cite{imran2014cloud}. This allowed us to eliminate biases.  
    \item\textbf{Automatic refactoring of smells:} Afterwards, we refactored the code from the smells. For each smell, we took a new copy of the smelly code and refactored to eliminate the effect of another refactoring. As a result, the projects were imported newly for each smell, refactored and executed. After refactoring, we repeat step 2 and record the CPU and memory consumption for the same workload and following the same method.
    \item\textbf{Logging the data:} We proceeded to log all the CPU and memory consumption in each experimental run. We tried to eliminate any external impact during each execution. We recorded the CPU and memory consumption during each CPU cycle and stored the data persistently.
    \item \textbf{Statistical operation:} Finally, we obtained the arithmetic mean of the CPU and memory usage. Afterward, we tabulated the results and compared whether the resource usage improved or deteriorated after refactoring the smells.
\end{itemize}
The next section highlights results obtained by following these prescribed steps.
\section{Results}
The obtained results during experimentation have been provided here. The number of detected smells is showcased which is followed by quantitative analysis of the data achieved during experimentation.

\subsection{Selected Software for Analysis} For this study, we identified and selected six open-source software which serve the users in the cloud. Among those, we analyzed 3,11,199 Lines of Code (LoC) and 6,249 class files. \textit{Zimbra} \cite{imran2019design} is an OS-independent emailing, and file sharing tool running in the cloud, popularly used by Dell, Rackspace, and Mozilla. \textit{OneDataShare} \cite{imran2019design} is a cloud-based data transfer tool that incorporates multiple transfer protocols, including DropBox and GoogleDrive. \textit{GraphHopper} is a GIS application provided as SaaS that incorporates spatial rules in hybrid mode.  \textit{Hadoop} is a software framework which was first proposed by Google to facilitate the analysis of large volume of data in the cloud \cite{ghazi2015hadoop}. \textit{Java Apache Mail Enterprise Server (JAMES)} consists of a modular architecture based on state of the art components which provides secure, stable, and end-to-end mail servers running on the JVM \cite{imran2019design}. \textit{JENA} is a platform provided by Apache for designing and building linked data applications, giving access to a variety of APIs for serialization, processing, storage, and transfer \cite{QualitasCorpus:APSEC:2010}. Table \ref{tab0} shows details related to the software selected for this study. Following is a description of the mechanism followed here to to obtain those software.
\begin{table}
\large
\centering
 \begin{tabular}{| c c c c c|} 
 \hline
 \scriptsize System & \scriptsize Activity &\scriptsize LoC &\scriptsize \scriptsize Commits &\scriptsize NoC\\ [0.3ex]
 \hline
 \scriptsize Zimbra &  \scriptsize 08/05-03/19 &\scriptsize 24,698 &\scriptsize 15,052 &\scriptsize 1,871 \\ 
 \scriptsize JGraph &\scriptsize 09/09-02/19 &\scriptsize 22,758 &\scriptsize 1,360 &\scriptsize 187 \\
 \scriptsize OnedataShare &\scriptsize 11/18-01/20 &\scriptsize 23,002 &\scriptsize 1,644 &\scriptsize 390 \\
 \scriptsize Hadoop &\scriptsize 12/11-03/20 &\scriptsize 14,2790 &\scriptsize 23,611 & \scriptsize 2,069 \\
 \scriptsize JENA &\scriptsize 06/10-05/19 &\scriptsize 70,948 &\scriptsize 8,287 &\scriptsize 1,392 \\
 \scriptsize JAMES &\scriptsize 06/04-08/19 &\scriptsize 27,003 &\scriptsize 9,314 &\scriptsize 340 \\ 
 \hline
\end{tabular}
\caption{List of selected systems for the study}
\label{tab0}
\end{table}

To analyze the impact of automatic refactoring of smells on resource usage, we decided to use open-source software that runs in the cloud as Software as a Service (SaaS) for multiple reasons. First, there are many software which are developed as SaaS for cloud since cloud computing has increased in popularity. So the availability of software was satisfied. Second, due to the high demand for the cloud, those software are heavily deployed by the industry. As shown in the previous paragraph, our selected software are used widely by top companies. Hence we believe it is important to analyze the impact of automatic smell refactoring on resource consumption since those companies will benefit from our research. Third, the authors of this paper felt that the software running in the cloud will contain a significant number of smells since those have complex code blocks that are written to ensure real-time user interactions and serving a large number of users from remote cloud data centers where they are hosted. Finally, one of the main challenges of SaaS is to optimize resource utilization, since provisioning extra resources in the cloud will require additional cost. Hence optimized use of cloud resources will result in cost reduction for the cloud service provider.

We obtained the source code of the software from Github during November 2019. The software source code obtained based on search in Github using the following keywords: \textit{cloud computing software, Software as a Service, pervasive computing,} and \textit{cloud data transfer}. From there, we selected the software sorted by popularity. To ensure that our selection of software is unbiased, we implement a set of checks which software shall satisfy to be selected for analysis. Those checks are described as follows:-

\begin{itemize}
    \item \textit{\textbf{Check 1 - Initial list of software:}} Initially, we conducted a preliminary search in Github to identify the realistic chance of obtaining software that runs in the cloud. This search was manual and it involved looking out for cloud computing SaaS. It helped us to identify the keywords for searching in Github and also laid the foundation for the following checks.
    
    \item \textit{\textbf{Check 2 - Script to automatically download the software:}} We developed a script that will parse through Github projects and automatically download the source codes from the master branch based on the keywords. It resulted in downloading 84 repositories from Github. We required to use automated scripts because cloning the 84 repositories manually would be time-consuming.
    
    \item \textit{\textbf{Check 3 - Refining the search to match tool requirements:}} Finally, we refined our search to be in line with the requirements of our selected automatic code smell detection and refactoring tools. Prior to that, we eliminated any source code not written in Java, which left us with a list of 21 repositories. Next, we decided to include source codes that can be compiled in Eclipse as \textit{JDeodrant} and \textit{JSparrow} both have plugins, which can be run using Eclipse. After this, we were left with 11 repositories. Finally, we considered repositories with at least 10,000 lines of code, since otherwise we could run into the risk of considering a prototype software which we aimed to avoid.
\end{itemize}


\subsection{Analysis of Results}
\begin{table*}[htbp]
\small
\caption{The impact of smell refactoring on percentage of CPU utilization}
\begin{center}
\begin{tabular}{|c|c|c|c|c|c|c|c|}
\hline
\textbf{Table}&\multicolumn{7}{|c|}{\textbf{Affect on CPU utilization after refactoring}} \\
\cline{2-8} 
\textbf{} & \textbf{\textit{cyclic dependency}}& \textbf{\textit{god method}}& \textbf{\textit{shotgun surgery}}& \textbf{\textit{spaghetti code}}& \textbf{\textit{feature envy}}& \textbf{\textit{type checking}}& \textbf{\textit{god class}}\\
\hline
Zimbra & -16.5\%& +3.5\% & -42.6\%& -27.6\%& +11.0\%& -33.7\%& +14.8\% \\
GraphHopper & -17.2\%&+14.6\% &-30.7\%&-0.7\%&+27.6\%&-10.4\%&+66.2\% \\
OneDataShare & -50.8\%&+35.8\% &-52.3\%&-27.4\%&+66.8\%&-53.6\%&+48.8\% \\
Hadoop & -4.7\%&+10.5\% &-14.4\%&-8.2\%&+20.1\%&-6.8\%&+17.4\% \\
JENA & -32.0\%&+14.4\%&-35.5\%&-10.4\%&+57.0\%&-2.8\%&+78.5\% \\
JAMES & -24.5\%&+3.7\% &-20.7\%&-2.7\%&+35.0\%&-3.3\%&+20.6\% \\
\hline

\end{tabular}
\label{tab1}
\end{center}
\end{table*}

\begin{table*}[htbp]
\small
\caption{The impact of smell refactoring on percentage of Memory utilization}
\begin{center}
\begin{tabular}{|c|c|c|c|c|c|c|c|}
\hline
\textbf{Table}&\multicolumn{7}{|c|}{\textbf{Affect on Memory utilization after refactoring}} \\
\cline{2-8} 
\textbf{} & \textbf{\textit{cyclic dependency}}& \textbf{\textit{god method}}& \textbf{\textit{shotgun surgery}}& \textbf{\textit{spaghetti code}}& \textbf{\textit{feature envy}}& \textbf{\textit{type checking}}& \textbf{\textit{god class}}\\
\hline
Zimbra & -60.8\%& +14.2\% & -71.2\%& -60.5\%& +12.6\%& -53.9\%& +6.2\% \\
GraphHopper & -55.6\%&+18.9\% &-57.1\%&-4.7\%&+66.7\%&-22.1\%&+54.8\% \\
OneDataShare & -33.2\%&+38.2\% &-31.4\%&-4.8\%8&+81.9\%&-13.6\%&+88.1\% \\
Hadoop & -7.5\%&+7.4\% &-12.8\%&-20.6\%&+5.4\%&-7.7\%&+17.3\% \\
JENA & -7.5\%&+48.6\%&-1.7\%&-14.2\%&+28.7\%&-2.9\%&+42.3\% \\
JAMES & -13.2\%&+0.1\% &-21.9\%&-17.8\%&+14.7\%&-5.0\%&+35.3\% \\
\hline

\end{tabular}
\label{tab2}
\end{center}
\end{table*}
We start with highlighting our findings related to CPU utilization. More specifically, we identify which code smells from the list will impact the CPU consumption of the analyzed Java software. Hence we address the research question \textit{"How does auto-refactoring of code smells in cloud-based software affect CPU utilization?"} Table \ref{tab1} identifies the change in CPU utilization after each smell is refactored. Each column highlights the percent change in CPU utilization after refactoring the smell specified in column heading. The negative sign in front of the numbers show decrease in resource use, whereas the positive sign show increase in resource usage.

After presenting the analysis of obtained results for CPU consumption, we analyze the results for change in memory usage after refactoring the smells. Hence, we aim to answer our second research question, \textit{"How does auto-refactoring of code smells in cloud-based software affect memory utilization?"} Table \ref{tab2} identifies the percent change in memory utilization after refactoring each of the smells. The values of memory usage were measured in MB. Following is an analysis of the obtained results.

As seen in Table \ref{tab1}, the refactoring of {\em cyclic dependency} smells improves CPU utilization for all six  software tested. The CPU utilization was reduced by 16.52\%, 17.22\%, 50.81\%, 4.67\%, 31.96\%, and 24.53\% respectively for \textit{Zimbra, GraphHopper, OneDataShare, Hadoop, JENA, and JAMES}. Cyclic dependencies result in direct and indirect dependencies between abstractions \cite{samarthyam2016refactoring}. The abstractions were tightly coupled between a large number of direct and indirect cyclic dependencies, so they resulted in a tangled design of the code. Further analysis showed that since the elements were not placed in the correct package, it resulted in cyclic dependencies between packages as well. Co-ordination between the packages became difficult which resulted in multiple calls to the same package. Hence, the existence of this smell resulted in higher CPU and memory usage. To eliminate the smell, JSparrow encapsulated all the packages involved in the chain and assigned it to a single team, which required less processing and could be loaded once into memory, thus reducing resource consumption in terms of both CPU and memory.

A common practice to eliminate the {\em god method} smells by refactoring tools is to detect and remove those by implementing {\em extract method} design \cite{perez2014analyzing}. This refactoring procedure will improve the quality and maintainability of the software while preserving correctness. However, our results in Tables \ref{tab1} and \ref{tab2} show that refactoring {\em god method} leads to an increase in resource utilization. 
The extra resource usage we found came from the higher number of message traffic obtained from architecture modification of this smell. Refactoring of this smell enables us to obtain a modular architecture of code where the elements have higher cohesion and lower coupling. Despite the benefits, this refactoring mechanism may not be ideal for cloud software as it might create harmful side effects in terms of sustainability \cite{perez2014analyzing}. Refactoring of {\em god method} smells yielded an increase in CPU utilization of \textit{3.50\%, 14.59\%, 35.79\%, 10.45\%, 14.36\%, and 3.69\% }for \textit{Zimbra, GraphHopper, OneDataShare, Hadoop, JENA, and JAMES}.

Removal of \textit{shotgun surgery} smells contributed to the reduction of resource usage by the software. As we know {\em shotgun surgery} smells occur when we try to modify a class which in turn makes multiple modifications to several different classes. For example, in \textit{OneDataShare}, removal of all instances of this smell resulted in 52.30\% of CPU and 31.43\% of memory utilization. We consulted the $diff$ in the commits of $OneDataShare$ and agreed that the smell was introduced due to the practice of overzealous and sudden changing of multiple components in different classes during development to incorporate new requirements without proper documentation. The refactoring tool used \textit{Inline Class} to transfer the diversified behaviors of one operation to one class. This was done for 39 occurrences of this smell. As a result, we obtained the improvement of resource utilization.

\textit{OneDataShare} was found to have improved CPU utilization by 27.35\% when the {\em spaghetti code} smells were refactored. On the other hand, \textit{Zimbra's} memory utilization improved by 60.47\% when this smell was removed from it. Once again taking $OneDataShare$ as  a case study, upon analysis of its source code using the refactoring tools, it was seen that the code used a large volume of $GOTO$ statements. Excessive use of $GOTO$ instead of well-articulated code design resulted in software that is convoluted and unmanageable \cite{ceccato2008goto}. At the same time, it causes the program to have methods scattered across many classes, which required more memory for file read and write operations. 

Eliminating {\em feature envy} causes increased resource utilization which is a threat to the sustainability of the software, as it would result in provisioning more resources for the same task, thus incurring more cost \cite{nongpong2015feature}. More specifically, CPU and memory consumption increased by \textit{66.84\%} and \textit{81.87\%} respectively when this smells was refactored in \textit{OneDataShare}.
Analysis of files in $OneDataShare$ showed that the method calls required access to specific classes each time it requested for certain operations, hence increasing inter-class communication, thus deteriorating memory and CPU usage. 

Refactoring the \textit{type checking} smell yielded CPU utilization improvement of 53.6\% of \textit{OneDataShare}. Also, notable improvement of memory utilization of 53.9\% was observed for \textit{Zimbra}. This is because type checking results in a large function which should be broken down into multiple smaller functions. Calling the large function numerous times will cause all its functionality to be executed even when it is not necessary, yielding high CPU and memory utilization.

The tested software started to consume an alarming quantity of memory after refactoring \textit{god class} smells. In general, the reason for a similar pattern of memory usage can be summarized to two main causes. The first is despite having a large volume of LoC and effective session cache management, the refactoring divided large classes into multiple sub-classes \cite{perez2014analyzing}. Hence, implementing \textit{extract method} on \textit{god classes} caused an increased number of calls the program has to make to perform its tasks, which increased the data volume stored in memory. Secondly, all the software have multiple developers contributing to it. So, from the perspective of community smells, there is a possible reason between these smells and developer viewpoint, which requires further research.

In summary, it is seen that all the software considered here suffered from increased CPU utilization after smells called \textit{god method, feature envy,} and \textit{god class} were refactored. The increase is \textit{3.5\%, 11.00\%, and 14.8\%} respectively for the three smells. Although the increase in CPU utilization may seem low, however, it is important to remember that when the software runs in the cloud, every CPU cycle is charged to the customer, hence increase in CPU utilization due to refactoring of software smells will lead to the cloud client incurring unnecessary extra cost. On the other hand, it is seen that refactoring the smells called \textit{cyclic dependency, shotgun surgery, spaghetti code,} and \textit{type checking} contributes to reduced CPU usage.

As seen in Table \ref{tab2}, memory consumption is seen to drastically increase after refactoring \textit{god class}. For \textit{Zimbra, GraphHopper, and OneDataShare} this increase in memory consumption is found to be \textit{6.2\%, 54.8\%, and 88.1\%}. Overall, similar observations are made for \textit{feature envy, and god method} smells as well for all six software. Like CPU, memory resource is also paid in the cloud, hence refactoring these smells will lead to additional cost as extra memory needs to be provisioned. We determine that the refactoring techniques adopted by \textit{JDeodrant} for \textit{god class,} and \textit{feature envy} smells, together with \textit{JSparrow} for \textit{god method} are not useful for the cloud-based applications since all the six software resulted in an increase of CPU utilization and memory utilization.

\section{Related Work}
Platform-specific code smells in High-Performance Computing applications were determined by Wang et al. \cite{wang2014platform}. AST based matching was used to determine smells present in an HPC software by comparing it with a dictionary of smells. The authors claimed that the removal of such smells would increase the speedup of the software when migrated to a new platform. The assumption was that certain code blocks perform well in terms of speedup in a given platform. However, the results show that certain smell detection and refactoring reduced the speedup, thus challenging the claims and showing the importance of further research in this area.

Oliveira et al. conducted an empirical study to evaluate nine context-aware Android apps to analyze the impact of automated refactoring of code smells on resource consumption \cite{oliveira2018empirical}. They studied three code smells, namely {\em god class, god method, and feature envy}. They found that for the three smells, resource utilization increases when they are refactored. Although their findings are useful, it is limited to the analysis of three code smells only. At the same time, the importance of analyzing the impact of automated code smell refactoring on cloud computing SaaS applications were not considered.

\section{Conclusion}
In this paper, we evaluated the impact of automatically refactoring seven code smells on resource usage of software running in the cloud platform. Obtained results highlight that the refactoring techniques adopted by \textit{JDeodrant} for \textit{god class,} and \textit{feature envy}, together with \textit{JSparrow} for \textit{god method} resulted in more message traffic which adversely affected CPU and memory usage. More specifically, cumulative increase of CPU and memory consumption for refactoring these three smells significantly higher as shown in Table \ref{tab1} and Table \ref{tab2}. Hence, there exists scope of further research to improve the automatic refactoring mechanisms of existing tools. Also, determining the correlation between refactoring multiple smells and resource consumption needs to be explored. Additionally, impact on resource usage after refactoring smells specific to the cloud should be studied.


\section*{Acknowledgment}
This project is in part sponsored by the National Science Foundation (NSF) under award numbers OAC-1724898 and OAC-1842054.

\bibliographystyle{IEEEtran}
\bibliography{SEKE2020}
\vspace{12pt}
\end{document}